\documentclass[preprint,prd,aps,showpacs,showkeys,nofootinbib]{revtex4}
\usepackage{graphicx}

\begin{document}

\title{Light neutralino dark matter in $U(1)_X$SSM}
\author{Shu-Min Zhao$^{1,2}$\footnote{zhaosm@hbu.edu.cn}, Guo-Zhu Ning$^{1,2}$\footnote{ninggz@hbu.edu.cn}, Jing-Jing Feng$^{1,2}$,
Hai-Bin Zhang$^{1,2}$, Tai-Fu Feng$^{1,2,3}$, Xing-Xing Dong$^{1,2}$}

\affiliation{$^1$ Department of Physics, Hebei University, Baoding 071002, China}
\affiliation{$^2$ Key Laboratory of High-precision Computation and Application of Quantum Field Theory of Hebei Province, Baoding 071002, China}
\affiliation{$^3$ Department of Physics, Chongqing University, Chongqing 401331, China}
\date{\today}

\begin{abstract}

  The $U(1)_X$ extension  of the minimal supersymmetric standard model(MSSM) is called as $U(1)_X$SSM with the local
 gauge group $SU(3)_C\times SU(2)_L \times U(1)_Y \times U(1)_X$. $U(1)_X$SSM has three singlet Higgs superfields beyond
 MSSM. In $U(1)_X$SSM, the mass matrix of neutralino is $8\times8$, whose lightest mass eigenstate possesses cold dark matter characteristic. Supposing the lightest neutralino as dark matter candidate, we study the relic density.  For dark matter scattering off nucleus, the cross sections including spin-independent and spin-dependent are both researched. In our numerical results, some parameter space can satisfy the constraints from the relic density and the experiments of dark matter direct detection.

\end{abstract}


\maketitle

\section{introduction}
There are several existences of dark matter in the universe, and dark matter contribution is more important
than the visible matter. The earliest and the most compelling evidences for dark matter are the luminous objects
that move faster than one expects\cite{rotation1}. The other evidences for the dark matter can be found in Refs.\cite{account1, account2}.
Besides the gravitational interaction, dark matter can take part in weak interaction\cite{OE3}. To keep the relic density
of dark matter, dark matter should be stable and live long. People have paid much attention to dark matter for many
years, but they have not known its mass and interaction property. The non-baryonic matter density
is $\Omega h^2=0.1186\pm0.0020$ \cite{pdg}, and the standard model(SM) can not explain this problem. It implies that there must be
new physics beyond the SM. From the present researches, axions, sterile neutrinos, weakly interacting massive
 particles (WIMPs)\cite{OE3, WIMP} etc. are dark matter candidates.

Considering the shortcoming of SM, physicists extend it and obtain a lot of extended models. In these new models,
the minimal supersymmetric standard model (MSSM)\cite{MSSM} is the favorite one, where the lightest neutralino can be dark matter candidate\cite{NDMMSSM}. Furthermore, MSSM is also extended by people, and its U(1) extensions are interesting\cite{UMSSM1,UMSSM2}. In this work, we extend the MSSM with the $U(1)_X$
gauge group\cite{U1X}. On the base of MSSM, we add three right-handed neutrinos and three
 singlet Higgs superfields $\hat{\eta},~\hat{\bar{\eta}},~\hat{S}$. The right-handed neutrinos can not only give tiny masses to
 light neutrinos but also produce the lightest scalar neutrino possessing dark matter character. Our $U(1)_X$ extension of MSSM is called as $U(1)_X$SSM\cite{Sarah,ZSMJHEP}, which relieves the so called
 little hierarchy problem that is in the MSSM.  The baryon number violating operators are avoided because the $U(1)_X$ gauge symmetry breaks spontaneously. So, the proton is stable.

 In $U(1)_X$SSM,  there are the terms $\mu\hat{H}_u\hat{H}_d$ and
$\lambda_H\hat{S}\hat{H}_u\hat{H}_d$.  $\hat{S}$ is the singlet Higgs superfield and possesses a non-zero VEV ($v_S/\sqrt{2}$).
Therefore, $U(1)_X$SSM has an effective $\mu_{eff}=\mu+\lambda_Hv_S/\sqrt{2}$, which will probably relieve the $\mu$ problem. As discussed in Ref.\cite{NMSSM},
 one can particularly put the $\mu$ term to zero by a redefinition $\frac{v_S}{\sqrt{2}}\rightarrow \frac{v_S}{\sqrt{2}}-\frac{\mu}{\lambda_H}$.
The singlet $S$ can improve the lightest CP-even Higgs mass at tree level. Then large loop corrections to the 125 GeV Higgs mass are
not necessary.  $S$ can also make the second light neutral CP-even Higgs heavy at TeV order. This can easily satisfy the constraints for heavy Higgs from
experiments(such as LHC). If we delete S, the second light neutral CP-even Higgs is light, whose mass varies from 150 GeV to 400 GeV. Considering the limit for the
heavy neutral CP-even Higgs, we should make the second light neutral CP-even Higgs heavier. The singlet $S$ can produce this effect.

In our previous work\cite{ZSMJHEP}, the lightest CP-even scalar neutrino is supposed as dark matter candidate in the framework of
$U(1)_X$SSM. Its relic density and the cross section
scattering from nucleus have been researched in detail. Some works of scalar neutrino dark matter can be found in Refs.\cite{Sneudark1,Sneudark2,CJJSN, SND}. Here, we study the lightest neutralino as dark matter candidate\cite{NDMEMSSM}.
In the base of the neutral bino ($\tilde{B}$), wino ($\tilde{W}^0$) and higgsinos ($\tilde{H}^0_d,\tilde{H}^0_u$),
the neutralino mass eigenstates in the MSSM has the parameters $\tan\beta,~ M_1,~M_2$ and $\mu$.
The lower limit on the lightest neutralino $\chi^0_1$ mass is about 46 GeV,
which can be derived from the Large Electron Positron (LEP) chargino mass limit\cite{DELP}. While, this limit increases to well above 100 GeV, in the constrained MSSM (cMSSM)\cite{cMSSM}. In pMSSM, the authors
research the lightest neutralino below 50 GeV satisfying the constraints from LHC and XENON100\cite{pMSSM}.
 Many people have studied the phenomenology of lightest neutralino in MSSM\cite{NDMMSSM} and there are a lot of works of
neutralino dark matter in several models. They enrich the dark matter research and give light to the direct research of dark matter.

After this introduction, some content of $U(1)_X$SSM is introduced in section II.
In section III, we suppose the lightest neutralino as a dark matter candidate and we study its relic density.
The direct detection of the lightest neutralino scattering off nuclei is reseached in section IV, which includes
both the spin-independent cross section and spin-dependent cross section.
The numerical results of the relic density and cross sections of dark matter scattering are all calculated in section V.
We give our discussion and conclusion in section VI. Some formulae are collected in the appendix.

\section{the $U(1)_X$SSM}

Extending the local gauge group from $SU(3)_C\otimes
SU(2)_L \otimes U(1)_Y$ to $SU(3)_C\otimes
SU(2)_L \otimes U(1)_Y\otimes U(1)_X$ and adding three Higgs singlets $\hat{\eta},~\hat{\bar{\eta}},~\hat{S}$, right-handed neutrinos $\hat{\nu}_i$ to MSSM, one can obtain $U(1)_X$SSM\cite{ZSMJHEP}. The introduction of right-handed neutrinos can explain the neutrino experiments.
The mass squared matrix of CP-even Higgs is $5\times5$, because the CP-even parts of
$\eta,~\bar{\eta},~ S$ mix with the neutral CP-even parts of
$H_u,~ H_d$.
We take into account the one loop corrections for the lightest CP-even Higgs with 125 GeV. The condition is similar for the CP-odd Higgs, whose mass squared matrix is also $5\times5$.
The sneutrinos are departed into CP-even sneutrinos and CP-odd sneutrinos, whose mass squared matrixes are both $6\times6$.
Here, we show the $U(1)_X$ charges of the MSSM superfields: $\hat{Q}_i (0)$, $\hat{u}^c_i(-\frac{1}{2})$, $\hat{d}^c_i(\frac{1}{2})$, $\hat{L}_i(0)$,
$\hat{e}^c_i(\frac{1}{2})$, $\hat{H}_u(\frac{1}{2})$, $\hat{H}_d(-\frac{1}{2})$.
In table I, the superfields beyond MSSM are collected in detail.
\begin{table}
\caption{ The $U(1)_X$SSM superfields beyond MSSM}
\begin{tabular}{|c|c|c|c|c|}
\hline
Superfields & $SU(3)_C$ & $SU(2)_L$ & $U(1)_Y$ & $U(1)_X$ \\
\hline
$\hat{\nu}_i$ & 1 & 1 & 0 & -$1/2$ \\
\hline
$\hat{\eta}$ & 1 & 1 & 0 & -1 \\
\hline
$\hat{\bar{\eta}}$ & 1 & 1 & 0 & 1\\
\hline
$\hat{S}$ & 1 & 1 & 0 & 0 \\
\hline
\end{tabular}
\label{quarks}
\end{table}

The concrete forms of the Higgs superfields are
\begin{eqnarray}
&&H_{u}=\left(\begin{array}{c}H_{u}^+\\{1\over\sqrt{2}}\Big(v_{u}+H_{u}^0+iP_{u}^0\Big)\end{array}\right),
~~~~~~
H_{d}=\left(\begin{array}{c}{1\over\sqrt{2}}\Big(v_{d}+H_{d}^0+iP_{d}^0\Big)\\H_{d}^-\end{array}\right),
\nonumber\\
&&\eta={1\over\sqrt{2}}\Big(v_{\eta}+\phi_{\eta}^0+iP_{\eta}^0\Big),~~~~~~~~~~~~~~~
\bar{\eta}={1\over\sqrt{2}}\Big(v_{\bar{\eta}}+\phi_{\bar{\eta}}^0+iP_{\bar{\eta}}^0\Big),\nonumber\\&&
\hspace{4.0cm}S={1\over\sqrt{2}}\Big(v_{S}+\phi_{S}^0+iP_{S}^0\Big).
\end{eqnarray}
$v_u$ and $v_d$ are the VEVs of the Higgs doublets $H_u$ and $H_d$. While, $v_\eta$,~ $v_{\bar\eta}$ and $v_S$ are the VEVs of
 the Higgs singlets $\eta$, $\bar{\eta}$ and $S$.
The angles $\beta$ and $\beta_\eta$ are defined as $\tan\beta=v_u/v_d$ and $\tan\beta_\eta=v_{\bar{\eta}}/v_{\eta}$.

The sneutrino fields
$\tilde{\nu}_L$ and $\tilde{\nu}_R$ read as
\begin{eqnarray}
\tilde{\nu}_L=\frac{1}{\sqrt{2}}\phi_l+\frac{i}{\sqrt{2}}\sigma_l,~~~~~~~~~~\tilde{\nu}_R=\frac{1}{\sqrt{2}}\phi_R+\frac{i}{\sqrt{2}}\sigma_R.
\end{eqnarray}

We show the superpotential and the soft breaking terms in $U(1)_X$SSM
\begin{eqnarray}
&&W=l_W\hat{S}+\mu\hat{H}_u\hat{H}_d+M_S\hat{S}\hat{S}-Y_d\hat{d}\hat{q}\hat{H}_d-Y_e\hat{e}\hat{l}\hat{H}_d+\lambda_H\hat{S}\hat{H}_u\hat{H}_d
\nonumber\\&&\hspace{1.0cm}+\lambda_C\hat{S}\hat{\eta}\hat{\bar{\eta}}+\frac{\kappa}{3}\hat{S}\hat{S}\hat{S}+Y_u\hat{u}\hat{q}\hat{H}_u+Y_X\hat{\nu}\hat{\bar{\eta}}\hat{\nu}
+Y_\nu\hat{\nu}\hat{l}\hat{H}_u+\mu_\eta\hat{\eta}\hat{\bar{\eta}}.
\nonumber\\
&&\mathcal{L}_{soft}=\mathcal{L}_{soft}^{MSSM}-B_SS^2-L_SS-\frac{T_\kappa}{3}S^3-T_{\lambda_C}S\eta\bar{\eta}
+\epsilon_{ij}T_{\lambda_H}SH_d^iH_u^j\nonumber\\&&
\hspace{1.0cm}-T_X^{IJ}\bar{\eta}\tilde{\nu}_R^{*I}\tilde{\nu}_R^{*J}
+\epsilon_{ij}T^{IJ}_{\nu}H_u^i\tilde{\nu}_R^{I*}\tilde{l}_j^J
-m_{\eta}^2|\eta|^2-m_{\bar{\eta}}^2|\bar{\eta}|^2-m_S^2S^2\nonumber\\&&
\hspace{1.0cm}-(m_{\tilde{\nu}_R}^2)^{IJ}\tilde{\nu}_R^{I*}\tilde{\nu}_R^{J}
-\frac{1}{2}\Big(M_X\lambda^2_{\tilde{X}}+2M_{BB^\prime}\lambda_{\tilde{B}}\lambda_{\tilde{X}}\Big)+h.c.
\end{eqnarray}
Here, $\mathcal{L}_{soft}^{MSSM}$ represents the soft breaking terms of MSSM.
Obviously, the $U(1)_X$SSM is more complicated than the MSSM.
 In our previous work, $Y^Y$ represents the $U(1)_Y$ charge and $Y^X$ denotes the $U(1)_X$ charge.
We have proven that $U(1)_X$SSM is anomaly free, and the details can be found in Ref.\cite{ZSMJHEP}.
In $U(1)_X$SSM, there are two Abelian groups $U(1)_Y$ and $U(1)_X$, which cause the gauge kinetic mixing.  This effect is the
characteristic beyond MSSM and it can also be induced through RGEs.

We write the covariant derivatives of $U(1)_X$SSM in the general form
\begin{eqnarray}
&&D_\mu=\partial_\mu-i\left(\begin{array}{cc}Y^Y,&Y^X\end{array}\right)
\left(\begin{array}{cc}g_{Y},&g_{{YX}}^{'}\\g_{{XY}}^{'},&g_{{X}}^{'}\end{array}\right)
\left(\begin{array}{c}A_{\mu}^{\prime Y} \\ A_{\mu}^{\prime X}\end{array}\right)\;,
\label{gauge1}
\end{eqnarray}
with $A_{\mu}^{\prime Y}$  ($A^{\prime X}_\mu$) denotes the gauge field of $U(1)_Y$ ($U(1)_X$).
Considering the fact that the two Abelian gauge groups are unbroken, we change
the basis through a correct matrix $R$  and redefine the $U(1)$ gauge fields
  \begin{eqnarray}
&&\left(\begin{array}{cc}g_{Y},&g_{{YX}}^{'}\\g_{{XY}}^{'},&g_{{X}}^{'}\end{array}\right)
R^T=\left(\begin{array}{cc}g_{1},&g_{{YX}}\\0,&g_{{X}}\end{array}\right)\;,~~~~~
R\left(\begin{array}{c}A_{\mu}^{\prime Y} \\ A_{\mu}^{\prime X}\end{array}\right)
=\left(\begin{array}{c}A_{\mu}^{Y} \\ A_{\mu}^{X}\end{array}\right)\;.
\label{gauge4}
\end{eqnarray}

Different from MSSM, the $U(1)_X$SSM gauge bosons $A^{X}_\mu,~A^Y_\mu$ and $V^3_\mu$ mix together at the tree level.
In the basis $(A^Y_\mu, V^3_\mu, A^{X}_\mu)$, the corresponding mass matrix reads as
\begin{eqnarray}
&&\left(\begin{array}{*{20}{c}}
\frac{1}{8}g_{1}^2 v^2 &~~~ -\frac{1}{8}g_{1}g_{2} v^2 & ~~~\frac{1}{8}g_{1}g_{{YX}} v^2 \\
-\frac{1}{8}g_{1}g_{2} v^2 &~~~ \frac{1}{8}g_{2}^2 v^2 & ~~~~-\frac{1}{8}g_{2}g_{{YX}} v^2\\
\frac{1}{8}g_{1}g_{{YX}} v^2 &~~~ -\frac{1}{8}g_{2}g_{{YX}} v^2 &~~~~ \frac{1}{8}g_{{YX}}^2 v^2+\frac{1}{8}g_{{X}}^2 \xi^2
\end{array}\right),\label{gauge matrix}
\end{eqnarray}
with $v^2=v_u^2+v_d^2$ and $\xi^2=v_\eta^2+v_{\bar{\eta}}^2$.
One can diagonalize the above mass matrix by an unitary matrix
including two mixing angles $\theta_{W}$ and $\theta_{W}'$. $\theta_{W}$ is Weinberg angle
and $\theta_{W}'$ is defined as
\begin{eqnarray}
\sin^2\theta_{W}'=\frac{1}{2}-\frac{(g_{{YX}}^2-g_{1}^2-g_{2}^2)v^2+
4g_{X}^2\xi^2}{2\sqrt{(g_{{YX}}^2+g_{1}^2+g_{2}^2)^2v^4+8g_{X}^2(g_{{YX}}^2-g_{1}^2-g_{2}^2)v^2\xi^2+16g_{X}^4\xi^4}}.
\end{eqnarray}

The lightest neutralino is supposed as dark matter candidate, and we obtain
the mass matrix of neutralino in the basis $(\lambda_{\tilde{B}}, \tilde{W}^0, \tilde{H}_d^0, \tilde{H}_u^0,
\lambda_{\tilde{X}}, \tilde{\eta}, \tilde{\bar{\eta}}, \tilde{s}) $. This is caused by the super partners of the added three Higgs
singlets and new gauge boson, which mix with the MSSM neutralino superfields.

\begin{eqnarray}
&&M_{\tilde{\chi}^0}=\left(
\begin{array}{cc}
\mathcal{ A} & \mathcal{B}  \\
 \mathcal{B}^T & \mathcal{C}
\end{array}
\right).
\end{eqnarray}
The concrete forms of $\mathcal{ A}$, $\mathcal{B}$ and $\mathcal{C}$ are
\begin{equation}
\mathcal{ A} = \left(
\begin{array}{cccc}
M_1 &0 &-\frac{g_1}{2}v_d &\frac{g_1}{2}v_u\\
0 &M_2 &\frac{1}{2} g_2 v_d  &-\frac{1}{2} g_2 v_u \\
-\frac{g_1}{2}v_d &\frac{1}{2} g_2 v_d  &0
&m_{\tilde{H}_u^0\tilde{H}_d^0} \\
\frac{g_1}{2}v_u &-\frac{1}{2} g_2 v_u  &m_{\tilde{H}_d^0\tilde{H}_u^0} &0
\end{array}
\right),~~~~~\label{neutralino}
  \mathcal{B} = \left(
\begin{array}{cccc}
{M}_{B B'} &0  &0  &0\\
 0 &0 &0 &0\\
m_{\lambda_{\tilde{X}}\tilde{H}_d^0} &0 &0 & - \frac{{\lambda}_{H} v_u}{\sqrt{2}}\\
m_{\lambda_{\tilde{X}}\tilde{H}_u^0} &0 &0 &- \frac{{\lambda}_{H} v_d}{\sqrt{2}}
\end{array}
\right),\label{neutralino}
 \end{equation}

 \begin{equation}
\mathcal{C}= \left(
\begin{array}{cccc}
{M}_{BL} &- g_{X} v_{\eta}  &g_{X} v_{\bar{\eta}}  &0\\
- g_{X} v_{\eta}  &0 &\frac{1}{\sqrt{2}} {\lambda}_{C} v_S+\mu_\eta  &\frac{1}{\sqrt{2}} {\lambda}_{C} v_{\bar{\eta}} \\
g_{X} v_{\bar{\eta}}  &\frac{1}{\sqrt{2}} {\lambda}_{C} v_S+\mu_\eta  &0 &\frac{1}{\sqrt{2}} {\lambda}_{C} v_{\eta} \\
 0 &\frac{1}{\sqrt{2}} {\lambda}_{C} v_{\bar{\eta}}
 &\frac{1}{\sqrt{2}} {\lambda}_{C} v_{\eta}  &m_{\tilde{s}\tilde{s}}\end{array}
\right),\label{neutralino}
 \end{equation}

\begin{eqnarray}
&& m_{\tilde{H}_d^0\tilde{H}_u^0} = - \frac{1}{\sqrt{2}} {\lambda}_{H} v_S  - \mu ,~~~~~~~
m_{\tilde{H}_d^0\lambda_{\tilde{X}}} = -\frac{1}{2} \Big(g_{Y X} + g_{X}\Big)v_d, \nonumber\\&&
m_{\tilde{H}_u^0\lambda_{\tilde{X}}} = \frac{1}{2} \Big(g_{Y X} + g_{X}\Big)v_u
 ,~~~~~~~~~~~~
m_{\tilde{s}\tilde{s}} = 2 M_S  + \sqrt{2} \kappa v_S.\label{neutralino1}
\end{eqnarray}
This matrix is diagonalized by $N$
\begin{equation}
N^* M_{\tilde{\chi}^0} N^{\dagger} = M^{diag}_{\tilde{\chi}^0}.
\end{equation}
It is too difficult to obtain exactly the analytic forms of the eigenvalues, eigenvectors and N for $M_{\tilde{\chi}^0}$.
With some supposition, we can deduce the lightest neutralino mass and eigenvector approximately.
Comparing with $\mathcal{A}$ and $\mathcal{C}$, the matrix $\mathcal{B}$ is very small.
In this condition, we can use $Z_N^T$ to simplify $M_{\tilde{\chi}^0}$ with matrix $\zeta$, whose elements are all small parameters of the order $\mathcal{B}/\mathcal{A}$.
\begin{eqnarray}
&&Z_N^T=\left(
\begin{array}{cc}
 1-\frac{1}{2}\zeta^T\zeta & -\zeta^T  \\
 \zeta & 1-\frac{1}{2}\zeta\zeta^T
\end{array}
\right),~~~~~~
Z_N^T.M_{\tilde{\chi}^0}.Z_N={\small\left(
\begin{array}{cc}
 \mathcal{K}_1 & 0  \\
 0 & \mathcal{K}_2
\end{array}
\right)},\nonumber\\&&
\mathcal{K}_1=\mathcal{A}-\frac{1}{2}(\zeta^T\zeta \mathcal{A}+\mathcal{A}\zeta^T\zeta)-\zeta^T\mathcal{B}^T-\mathcal{B}\zeta+\zeta^T\mathcal{C}\zeta ,
\nonumber\\&& \mathcal{K}_2=\zeta \mathcal{A}\zeta^T+\mathcal{B}^T\zeta^T+\zeta \mathcal{B} +\mathcal{C} -\frac{1}{2}(\zeta\zeta^T \mathcal{C}+\mathcal{C}\zeta\zeta^T) .
\end{eqnarray}

 $\zeta^T$ can be calculated from the equation  $\mathcal{A}\zeta^T+\mathcal{B}-\zeta^T\mathcal{C}=0$.
 If we take the simplest approximation, it is
  \begin{eqnarray}
Z_N^T.M_{\tilde{\chi}^0}.Z_N
\sim\left(
\begin{array}{cc}
 \mathcal{A} & 0  \\
 0 & \mathcal{C}
\end{array}
\right).
\end{eqnarray}
In this work, we suppose the lightest neutralino $m_{\chi^0_1}$ is different from the MSSM condition, that is to say $m_{\chi^0_1}$ dominantly comes from the
matrix $\mathcal{C}$ and MSSM neutralinos in the matrix $\mathcal{A}$ are heavy.  Therefore, we calculate the
mass eigenstates of $\mathcal{C}$, which is tedious to solve
 the common quartic equation with one unknown quantity. Considering the constraint that $\tan\beta_\eta$ is near 1, we use
 $s_m=v_{\bar{\eta}}-v_\eta$ with the relation $v_\eta \gg s_m$. So, $\mathcal{C}$ turns to

\begin{eqnarray}
\mathcal{C}=\left(
\begin{array}{cccc}
 M_{BL} & -g_X v_\eta & g_X v_\eta+g_X s_m & 0 \\
  -g_X v_\eta & 0 & \lambda_C^\prime v_S+\mu_\eta &\lambda_C^\prime v_\eta +\lambda_C^\prime s_m
   \\
 g_X v_\eta+g_X s_m & \lambda_C^\prime v_S+\mu_\eta & 0 & \lambda_C^\prime v_\eta \\
  0 & \lambda_C^\prime v_\eta +\lambda_C^\prime s_m  & \lambda_C^\prime v_\eta & m_{\tilde{s}\tilde{s}}
\end{array}
\right),
\end{eqnarray}
with $\lambda_C^\prime=\lambda_C/\sqrt{2}$.

The eigenvalues of $\mathcal{C}$ are deduced to the leading order according to the small parameter $s_m$.
\begin{eqnarray}
&&m_{\chi^0_{a}}^{(0)}= \frac{1}{2} \left(M_{BL}-\lambda_C^\prime
   v_S-\mu_\eta+\sqrt{8 g_X^2
   v_\eta^2+(\lambda_C^\prime v_S+\mu_\eta+M_{BL})^2}\right),
   \nonumber\\&&
   m_{\chi^0_{b}}^{(0)}= \frac{1}{2} \left(-M_{BL}+\lambda_C^\prime
   v_S+\mu_\eta+\sqrt{8 g_X^2
   v_\eta^2+(\lambda_C^\prime v_S+\mu_\eta+M_{BL})^2}\right),\nonumber\\&&
m_{\chi^0_{c}}^{(0)}= \frac{1}{2} \left(\lambda_C^\prime
   v_S+\mu_\eta+m_{\tilde{s}\tilde{s}}+\sqrt{8 (\lambda_C^\prime)^2
   v_\eta^2+(m_{\tilde{s}\tilde{s}}-\lambda_C^\prime v_S-\mu_\eta)^2}\right),
   \nonumber\\&&
m_{\chi^0_{d}}^{(0)}= \frac{1}{2} \left(\lambda_C^\prime
   v_S+\mu_\eta+m_{\tilde{s}\tilde{s}}-\sqrt{8 (\lambda_C^\prime)^2
   v_\eta^2+(m_{\tilde{s}\tilde{s}}-\lambda_C^\prime v_S-\mu_\eta)^2}\right).
   \end{eqnarray}
We take $M_{BL}$ and $m_{\tilde{s}\tilde{s}}$ are both positive parameters. To satisfy the constraint from $m_{Z^\prime}$,
$v_\eta$ is large and bigger than $v_S$. So it is easy to see that $m_{\chi^0_{a}}$ and $m_{\chi^0_{c}}$ are large values.
Using $m_{\tilde{s}\tilde{s}}\gg M_{BL}$ and $\mu_\eta$, $m_{\chi^0_{d}}$ is smaller than $m_{\chi^0_{b}}$, so $m_{\chi^0_{d}}$ is the lightest
neutralino mass $m_{\chi^0_{1}}$. For the lightest neutralino mass, we consider the correction at the order $s_m$.
   \begin{eqnarray}
&&m_{\chi^0_{d}}^{(1)}=-\frac{2 (\lambda_C^\prime)^2 s_m v_\eta}{\sqrt{8  (\lambda_C^\prime)^2
   v_\eta^2+(m_{\tilde{s}\tilde{s}}-\lambda_C^\prime v_S-\mu_\eta)^2}}.
\end{eqnarray}
At the leading order, the eigenvector of $m_{\chi^0_{d}}$ is
\begin{eqnarray}
&&V^{(0)}_{\chi^0_{d}}= \frac{1}{\sqrt{2+a_0^2}}\Big(0,1,1,a_0\Big)\sim \Big(0,\frac{1}{\sqrt{2}},\frac{1}{\sqrt{2}},0\Big),
\nonumber\\&&
a_0=\frac{4 (\lambda_C^\prime)v_\eta}{\lambda_C^\prime
   v_S+\mu_\eta-m_{\tilde{s}\tilde{s}}-\sqrt{8(\lambda_C^\prime)^2
   v_\eta^2+(m_{\tilde{s}\tilde{s}}-\lambda_C^\prime v_S-\mu_\eta)^2}}.
\end{eqnarray}
Here, $a_0$ is a small parameter. The $s_m$ correction to eigenvector $V^{(0)}_{\chi^0_{d}}$ is
$V^{(1)}_{\chi^0_{d}}$
\begin{eqnarray}
&&V^{(1)}_{\chi^0_{d}}=\frac{s_m}{v_\eta}\frac{1}{\sqrt{b_1^2+c_1^2}}\Big( 0,~0,~b_1,~ c_1\Big),\nonumber\\&&
b_1=\frac{\sqrt{8 (\lambda_C^\prime)^2 v_\eta^2+m_{\tilde{s}\tilde{s}}^2}+m_{\tilde{s}\tilde{s}}}{4 \lambda_C^\prime
   v_\eta},~~~
c_1=\frac{1}{2} \Big(\frac{m_{\tilde{s}\tilde{s}}}{\sqrt{8 (\lambda_C^\prime)^2
   v_\eta^2+m_{\tilde{s}\tilde{s}}^2}}-1\Big).
\end{eqnarray}
$c_1$ is much smaller than $b_1$, then $V^{(1)}_{\chi^0_{d}}$ can be simplified as
\begin{eqnarray}
V^{(1)}_{\chi^0_{d}}\sim\frac{s_m}{v_\eta}\Big( 0,~0,~1,~ 0\Big).
\end{eqnarray}
In the whole, the eigenvector of the lightest neutralino is dominatly composed by the linear combination
of $\tilde{\eta}$ and $\tilde{\bar{\eta}}$.

In the MSSM, the lightest CP-even Higgs mass at tree level is no more than 90 GeV, and
 the loop corrections to the lightest CP-even Higgs mass  can be large.
 Including the leading-log radiative corrections from stop and top particles \cite{higgsOL}, we write the mass of the
  lightest CP-even Higgs boson in the following form
\begin{eqnarray}
&&m_h=\sqrt{(m_{h_1}^0)^2+\Delta m_h^2}.\label{higg mass}
\end{eqnarray}
Here, $m_{h_1}^0$ represents the lightest Higgs boson mass at tree level and $\Delta m_h^2$ is shown analytically
\begin{eqnarray}
&&\Delta m_h^2=\frac{3m_t^4}{2\pi v^2}\Big[\Big(\tilde{t}+\frac{1}{2}+\tilde{X}_t\Big)+\frac{1}{16\pi^2}\Big(\frac{3m_t^2}{2v^2}-32\pi\alpha_3\Big)\Big(\tilde{t}^2
+\tilde{X}_t \tilde{t}\Big)\Big],\nonumber\\
&&\tilde{t}=\log\frac{M_{\tilde{T}}^2}{m_t^2},\qquad\;\tilde{X}_t=\frac{2\tilde{A}_t^2}
{M_{\tilde{T}}^2}\Big(1-\frac{\tilde{A}_t^2}{12M_{\tilde{T}}^2}\Big),\label{higgs corrections}
\end{eqnarray}
with $\alpha_3$ denoting the strong coupling constant. The parameter $\tilde{A}_t$ is $\tilde{A}_t=A_t-\mu \cot\beta$
with $A_t$ representing the trilinear Higgs stop coupling. $M_{\tilde{T}}=\sqrt{m_{\tilde t_1}m_{\tilde t_2}}$
and $m_{\tilde t_{1,2}}$ are the stop masses.
To save space in the text, other used couplings are collected in the appendix.

\section {relic density}

Supposing the lightest neutralino($\chi^0_1$) as dark matter candidate, we calculate the relic density. The constraint of dark matter relic density is severe, and the concrete value is  $\Omega_D h^2=0.1186\pm 0.0020$ \cite{pdg}.
The $\chi^0_1$ number density $n_{\chi^0_1}$ should satisfy the Boltzmann  equation \cite{boltzmann,XFBO1}
\begin{eqnarray}
\frac{d n_{\chi^0_1}}{dt}=-3Hn_{\chi^0_1}-\langle\sigma v\rangle_{SA}(n^2_{\chi^0_1}-n^2_{\chi^0_1 eq})
-\langle\sigma v\rangle_{CA}(n_{\chi^0_1}n_\phi-n_{\chi^0_1 eq}n_{\phi eq}).
\end{eqnarray}

For $\chi^0_1$,  we take into account self-annihilation and co-annihilation with another particle $\phi$.
At the temperature $T_F$, the
 annihilation rate of $\chi^0_1$ is approximately equal to the Hubble expansion rate, and the lightest neutralino  freezes out.
 We suppose $\chi^0_1$ is the lightest SUSY particle and $m_\phi$ is larger than $m_{\chi^0_1}$. The relevant
 formulae are \cite{ExpYD}
\begin{eqnarray}
&&\langle\sigma v\rangle_{SA}n_{\chi^0_1}+\langle\sigma v\rangle_{CA}n_{\phi}\sim H(T_F),\nonumber\\&&
n_\phi=\Big(\frac{m_\phi}{m_{\chi^0_1}}\Big)^{3/2}\texttt{Exp}[(m_{\chi^0_1}-m_\phi)/T]n_{\chi^0_1}\nonumber\\&&
\Big[\langle\sigma v\rangle_{SA}+\langle\sigma v\rangle_{CA}
\Big(\frac{m_\phi}{m_{\chi^0_1}}\Big)^{3/2}\texttt{Exp}[(m_{\chi^0_1}-m_\phi)/T]\Big]n_{\chi^0_1}\sim H(T_F).
\end{eqnarray}

After we study the self-annihilation cross section $\sigma(\chi^0_1\chi^0_1 \rightarrow$ anything) and
co-annihilation cross section $\sigma(\chi^0_1 \phi \rightarrow$ anything), $\langle\sigma v\rangle_{SA}$ and $\langle\sigma v\rangle_{CA}$
are gotten. The annihilation results can be written as
$\sigma v_{rel}=a+bv_{rel}^2$ in the mass center frame. Here, $v_{rel}$ is
the relative velocity of the two particles in the initial states. Using the following formula,
we can approximately calculate the freeze-out temperature ($T_F$)\cite{NDMMSSM,XFCW,XFBO1}
\begin{eqnarray}
&&x_F=\frac{m_{\chi_1^0}}{T_F}\simeq\ln[\frac{0.076M_{Pl}m_{\chi_1^0}(a+6b/x_F)}{\sqrt{g_*x_F}}],
\end{eqnarray}
with $M_{Pl}$ denoting the Planck mass.

The relativistic degrees of freedom with mass less than $T_F$ is represented by $g_*$.
The cold non-baryonic dark matter density is simplified in the following form\cite{rotation1,XFBO1,zhaosm}.
\begin{eqnarray}
\Omega_D h^2\simeq \frac{1.07\times10^9 x_F}{\sqrt{g_*}M_{PL}(a+3b/x_F)~\rm{GeV} }\;.
\end{eqnarray}

It is well known that, the self-annihilation processes are dominant in general condition. We show the researched concrete
self-annihilation processes:
 $\chi^0_1+\chi^0_1\rightarrow A+B$, A and B represent final states  $(Z,~h),~(W,~W),~(Z,~Z),~(h,~h),~(\bar{u}_i, u_i),
~(\bar{d}_i,~ d_i),~(\bar{l}_i,~ l_i),~(\bar{\nu}_i,~ \nu_i)$.  Here $i=1,~2,~3$ and $h$ represents the lightest CP-even Higgs.
 The neutrinos in final state are just three light neutrinos not including heavy neutrinos.

For co-annihilation processes, if the mass of another particle is almost equal to the mass of $\chi^0_1$,
they give considerable contributions to the annihilation cross section.

a. The lightest neutralino $\chi^0_1$ annihilates with heavier neutralinos $\chi^0_k(k=2\dots8)$, whose final states
are same as those produced by self-annihilation processes.

b. $\chi^0_1+\chi^{-}\rightarrow \{(\nu,l^-),(\bar{u},d),(W^-,Z),(W^-,\gamma),(W^-,h^0)\} $.
The corresponding processes are obtained by the charge conjugate transformation.

c. $\chi^0_1+\tilde{L}^-\rightarrow \{(\gamma,l^-),(Z,l^-),(h^0,l^-)\} $. Similar as the condition b, condition c also
has charge conjugate processes.

d. $\chi^0_1+\tilde{\nu}^R(\tilde{\nu}^I)\rightarrow \Big\{(\nu,Z), (l^-,W^+), (l^+,W^-)$\Big\}.

  The co-annihilation between $\chi_1^0$ and  scalar quarks($\tilde{U},~\tilde{D}$) are neglected, because scalar quark masses
  are very heavy and much larger than $m_{\chi^0_1}$.

\section{direct detection}

The experiment constraints for the direct detection of dark matter become strict more and more.
The lightest neutralino scatters off nucleus, and the process is  $\chi^0_1+q\rightarrow \chi^0_1+ q$.
The exchanged particles can be CP-even Higgs $H^0_j$, CP-odd Higgs $A^0_j$, gauge bosons $Z,~Z^\prime$.
For the CP-odd Higgs $A^0_j$ contribution, there are two suppression factors: 1. The Yukawa coupling $Y_q$
of light quark; 2. The operators $\bar{\chi}^0_1\chi^0_1\bar{q}\gamma_5q$ and $\bar{\chi}^0_1\gamma_5\chi^0_1\bar{q}\gamma_5q$
are suppressed by the factors $q^2$ and $q^4$ respectively\cite{LJandHe}. Therefore, we neglect the CP-odd Higgs contribution.
Because neutralino is Majorana particle, the operator $\bar{\chi}^0_1\gamma_\mu\chi^0_1\bar{q}\gamma^\mu q$ disappears.
The dominant operators at quark level are $\bar{\chi}^0_1\chi^0_1\bar{q}q$ and $\bar{\chi}^0_1\gamma_\mu\gamma_5\chi^0_1\bar{q}\gamma^\mu\gamma_5 q$ obtained from CP-even Higgs and vector bosons $Z,~Z^\prime$ contributions\cite{LJandHe}.

The quark level operators should be converted to the effective nucleus operators.
To convert the operator $\bar{\chi}^0_1\chi^0_1\bar{q}q$, we use the following formulae\cite{LJandHe}
\begin{eqnarray}
&&a_qm_q\bar{q}q\rightarrow f_Nm_N\bar{N}N,~~~~~~~~~~~~~~~~~~~~~~~~~\langle N|m_q \bar{q}q|N\rangle=m_N f_{Tq}^{(N)},
\nonumber\\&&f_N=\sum_{q=u,d,s}f_{Tq}^{(N)}a_q+\frac{2}{27}f_{TG}^{(N)}\sum_{q=c,b,t}a_q,
~~~~~~f_{TG}^{(N)}=1-\sum_{q=u,d,s}f_{Tq}^{(N)}.
\end{eqnarray}

Integrating out heavy quark loops, the coupling to gluons is induced, which is included in
$f_N$. We show the concrete values of the parameters $f_{Tq}^{(N)}$ \cite{DarkSUSY1},
\begin{eqnarray}
&&f^{(p)}_{Tu}=0.0153,~~~f^{(p)}_{Td}=0.0191,~~~f^{(p)}_{Ts}=0.0447, \nonumber\\&&
f^{(n)}_{Tu}=0.0110,~~~f^{(n)}_{Td}=0.0273,~~~f^{(n)}_{Ts}=0.0447.
\end{eqnarray}

$\bar{\chi}^0_1\gamma_\mu\gamma_5\chi^0_1\bar{q}\gamma^\mu\gamma_5 q$ is a spin-dependent operator, which is converted to the
effective nucleus operator with the following formulae.
\begin{eqnarray}
 &&d_q\bar{q}\gamma^\mu\gamma_5 q\rightarrow a_N\bar{N}s^{(N)}_\mu N,~~~\langle N | \bar{q}\gamma_\mu\gamma_5q|N\rangle=
 s^{(N)}_\mu\Delta q^{(N)}, ~~~ a_N=\sum_{u,d,s}d_q\Delta q^{(N)},
\end{eqnarray}
with $s^{(N)}_\mu$ denoting the spin of nucleus. In the numerical calculation, we use the parameters of DarkSUSY
\begin{eqnarray}
\Delta u^{(p)}=\Delta d^{(n)}=0.77,~~~\Delta d^{(p)}=\Delta u^{(n)}=-0.47,~~~\Delta s^{(p)}=\Delta s^{(p)}=-0.15.
\end{eqnarray}

For the spin-independent operator $\bar{\chi}^0_1\chi^0_1\bar{q}q$, the scattering cross section reads as\cite{LJandHe}
\begin{eqnarray}
\sigma=\frac{1}{\pi}\hat{\mu}^2[Z_pf_p+(A-Z_p)f_n]^2,
\end{eqnarray}
with $Z_p$ denoting the number of proton, and $A$ representing the number of atom.

The scattering cross section for the spin-dependent operator $\bar{\chi}^0_1\gamma_\mu\gamma_5\chi^0_1\bar{q}\gamma^\mu\gamma_5 q$
is shown as \cite{LJandHe}
\begin{eqnarray}
\sigma=\frac{16}{\pi}\hat{\mu}^2a_N^2J_N(J_N+1),
\end{eqnarray}
with $J_N$ is the number of angular momentum for the nucleus.
The corresponding formula for one nucleon is
\begin{eqnarray}
\sigma=\frac{12}{\pi}\hat{\mu}^2a_N^2.
\end{eqnarray}

\section{numerical results}
To study the numerical results,  we should take into account the experimental constraints. One strict constraint from experiment
is the mass(125 GeV)\cite{mh01} of the lightest CP-even Higgs. $Z'$ boson mass constraint is also important.  The mass bounds for $M_{Z'}$ from LHC
are more severe than the limits from the low energy data. In the Sequential Standard Model, the lower mass limit of
$Z^\prime _{SSM}$ is 4.5 TeV at $95\% $ confidence level(CL).
The Lower mass limits of the $Z^\prime$ boson in the left-right symmetric model and the (B-L) model \cite{ATLAS2016}
are respectively 4.1 TeV and 4.2 TeV.
The upper bound on the ratio between $M_{Z^\prime}$ and its gauge
coupling is $M_{Z^\prime}/g_X\geq6$ TeV at 99\% CL\cite{ZPG1,ZPG2}.
  Considering the LHC experimental data, $\tan \beta_\eta$ should be smaller than 1.5 \cite{TanBP}.
  We take into account the above constraints and  choose the parameters to satisfy the relation $M_{Z^{\prime}}> 4.5$ TeV\cite{ZSMJHEP}.

Here, we also add other experiment limits.  The considered mass limits for the particles beyond SM are\cite{pdg}:
 1 the mass limits for heavy neutral Higgs $(H^0,A^0)$ and charged Higgs ($H^\pm$); 2 the mass limits for neutralino, chargino, sneutrino, scalar charged lepton, squark.
 The decays of the lightest CP even Higgs ($m_{h^0}=125$ GeV) such as $h^0\rightarrow \gamma+\gamma$,
  $h^0\rightarrow Z+Z$ and $h^0\rightarrow W+W$ are considered. The constraint from $B\rightarrow X_s+\gamma$ is also taken into account. With new experiment data of muon g-2 from the Fermion National Accelerator Laboratory(FNAL)\cite{muong2},  the deviation between experiment and SM prediction
  is $\Delta a_\mu=a^{exp}_\mu-a^{SM}_\mu=251(59)\times 10^{-11}$ and increases to 4.2$\sigma$.
  We study muon g-2 in $U(1)_XSSM $ in the previous work\cite{SULH}, and consider this limit here.

Therefore, we use the following parameters
\begin{eqnarray}
&&M_S =2.7 ~{\rm TeV},~
T_{\kappa} =1.6~ {\rm TeV}, ~g_{YX}=0.2,~g_X=0.3,~\lambda_C = -0.08,~\lambda_H = 0.1,\nonumber\\&&
\upsilon_{\eta} = 15.5\times\cos\beta_\eta ~{\rm TeV},~
\upsilon_{\bar{\eta}} = 15.5\times\sin\beta_\eta~{\rm TeV},
~Y_{X11} =Y_{X22} = 0.5,~Y_{X33} =0.4,
\nonumber\\&&
~T_{\lambda_H} = 0.3~{\rm TeV},~
T_{X11} =T_{X22} = T_{X33} = -1~{\rm TeV}, ~
T_{e11} =T_{e22} = T_{e33} = -3~{\rm TeV},\nonumber\\&&
T_{\lambda_C} =- 0.1~{\rm TeV},\mu_\eta=10~{\rm GeV},~M^2_{U11}=M^2_{U22}=10~{\rm TeV}^2,~M^2_{Q33}=M^2_{U33}=3.5~{\rm TeV}^2,\nonumber\\&&
 l_W = 4~{\rm TeV}^2,~
M^2_{\nu11} = M^2_{\nu22}=M^2_{\nu33}=0.5~{\rm TeV}^2,~T_{u11}=T_{u22}=T_{u33}=-2~{\rm TeV},~\kappa=1,\nonumber\\&&T_{d12}= T_{d21}=0.2~{\rm TeV},~
 M^2_{L11}= M^2_{L22} = M^2_{L33} =1.9~{\rm TeV}^2,~B_{\mu} = B_S=m_S^2=1~{\rm TeV}^2,\nonumber
\\&& M^2_{E11}= M^2_{E22} = M^2_{E33} =3.54 ~{\rm TeV}^2,~ \tan\beta_\eta=0.8,
~T_{\nu11} = T_{\nu22}=T_{\nu33}=0.5 ~{\rm TeV}.
\end{eqnarray}
To simplify the numerical discussion, most of the parameters $T_\nu,~T_X, ~T_u $ etc. are supposed as diagonal matrices and we
use the supposition
\begin{eqnarray}
M^2_{D11}=M^2_{D22}= M^2_{D33}=M_D^2,~~~~~~T_{d11}= T_{d22}= T_{d33}=T_d.
\end{eqnarray}

\subsection{The relic density of neutralino dark matter}

With the supposition that the lightest neutralino $\chi^0_1$ is the lightest SUSY particle(LSP),
we research the relic density of $\chi^0_1$.
In this subsection, we adopt the parameters as
$ M^2_{Q11}=M^2_{Q22}=M^2_{D}=10 {\rm TeV}^2$ and $T_d=1 {\rm TeV}$.

\begin{figure}[h]
\setlength{\unitlength}{1mm}
\centering
\includegraphics[width=3.6in]{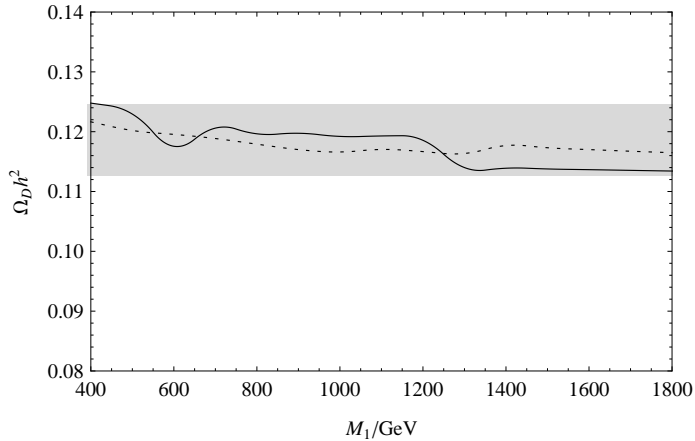}
\caption[]{ The relic density $\Omega_D h^2$ versus $M_1$ is plotted by the solid line (dotted line) with $M_2=1(2)$TeV.\label{M1}}
\end{figure}

$M_1$ is the mass of $U(1)_Y$ gaugino and
appears in the neutralino mass matrix. Therefore, $M_1$ can affect neutralino masses and mixing to some extent.
In the Fig.\ref{M1}, we plot the relic density in the banded gray area with $\pm3\sigma$ sensitivity.
 The relic density versus $M_1$ is represented by solid line($M_2=1 {\rm TeV}$) and dotted line($M_2=2 {\rm TeV}$)
with the parameters $ M_{BL}=1 {\rm TeV},~\tan\beta =9,~\mu=0.5 {\rm TeV}$ and $ M_{BB^\prime}=0.4 {\rm TeV}$.
 It is obvious that both the solid line and the dotted line versus $M_1$ vary weakly in the region $[400,~1800]$ GeV.
 The both lines are in the $\pm3\sigma$ band. At the point $M_1=1200$ GeV, the relic density is very near its central value,
 which  can well satisfy the experiment constraint. Generally speaking, the two lines are very near.
 With the used parameters, the mass of the lightest neutralino $\chi^0_1$ is around 302 GeV, and the other SUSY particles are all much heavier than $\chi^0_1$. So, the self-annihilation processes are dominant. That is to say, the contributions from the co-annihilation
 processes are tiny. Because the masses of exchanged virtual particles are not near $2*m_{\chi^0_1}$, the resonance annihilation
affecting the relic density strongly can not take place.

In this parameter space, the masses of some SUSY particles that can  co-annihilate with the lightest neutralino are collected here: the second light neutralino mass $m_{\chi_2^0}\sim800{\rm GeV}$, the lightest scalar neutrino mass (CP-even and CP-odd) $m_{\tilde{\nu}}\sim 1600{\rm GeV}$, the lightest scalar lepton mass $m_{\tilde{L}_1}\sim 880 {\rm GeV}$, the lightest chargino mass $m_{\chi^\pm_1}\sim 780 {\rm GeV}$,
the lightest scalar quark mass $m_{\tilde{q}_1}\sim1800{\rm GeV}$.

\begin{figure}[h]
\setlength{\unitlength}{1mm}
\centering
\hspace{1.3cm}\includegraphics[width=3.6in]{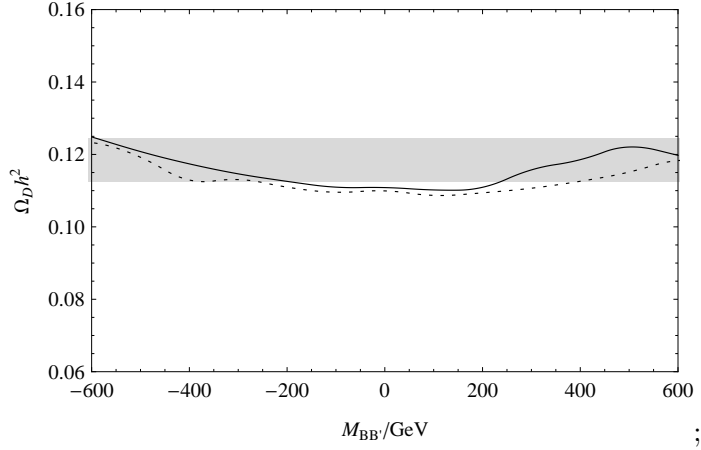};
\vspace{0cm}
\caption[]{The relic density $\Omega_D h^2$ versus $M_{BB^\prime}$ is plotted by the solid line (dotted line) $\mu$=0.5(0.4)TeV.}\label{MBBp}
\end{figure}

$M_{BB^\prime}$ is the mass of the $U(1)_Y$ and $U(1)_X$ gaugino mixing and presents in the mass matrix of neutralino, which can affect the relic density through the mixing matrix.
Here, we use the parameters as $M_1 =1.2 {\rm TeV},~M_2 =1 {\rm TeV}, ~M_{BL}=1 {\rm TeV},~\tan\beta =9$.
In the Fig.\ref{MBBp}, the numerical results of the relic density versus $M_{BB^\prime}$ are shown by the solid line($\mu=0.5$ TeV) and dotted line($\mu=0.4$ TeV) respectively. The solid line is above the dotted line.  When $M_{BB^\prime}$ is near zero, $\Omega_Dh^2$ can not satisfy the experimental constraint. The numerical results of $\Omega_Dh^2$ corresponding to $M_{BB^\prime}$ regions [-500, -300]GeV and [300, 500] GeV are better.

\begin{figure}[h]
\setlength{\unitlength}{1mm}
\centering
\hspace{1.3cm}\includegraphics[width=3.6in]{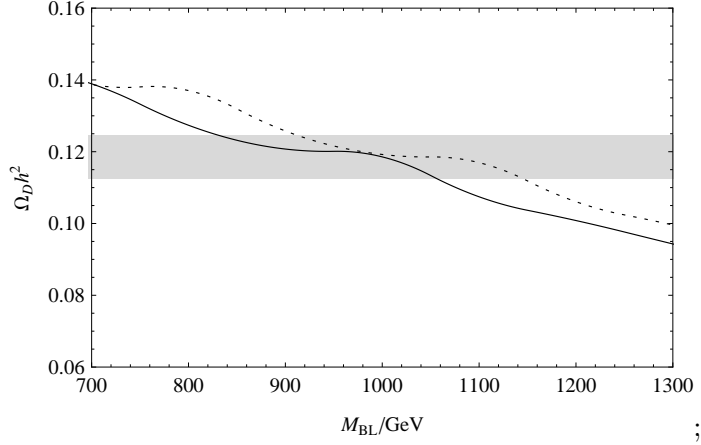};
\vspace{0cm}
\caption[]{The relic density $\Omega_D h^2$ versus $M_{BL}$ is plotted by the solid line (dotted line) $\tan\beta$=9(5).}\label{MBL}
\end{figure}

$M_{BL}$ is the mass of the new gaugino, and it has influence on the mass matrix of neutralino.
Therefore, $M_{BL}$ can have considerable effect on the relic density. With the parameters
$M_1 =1.2 {\rm TeV},~M_2 =1 {\rm TeV},~\mu=0.5{\rm TeV},~ M_{BB^\prime}=0.4{\rm TeV}$, we plot the relic density
versus $M_{BL}$ in the Fig.\ref{MBL}. The solid line corresponds to $\tan\beta=9$, and the dotted line corresponds to
$\tan\beta=5$. Both the solid line and the dotted line become small with the increasing $M_{BL}$, and they possess similar behavior.
For the both lines, the best point is around $M_{BL}=1000$ GeV.
As  $M_{BL}<800$ GeV or $M_{BL}>1150$ GeV, the obtained numerical results of $\Omega_Dh^2$
exceed the experimental data.

\begin{figure}[h]
\setlength{\unitlength}{1mm}
\centering
\hspace{1.3cm}\includegraphics[width=3.6in]{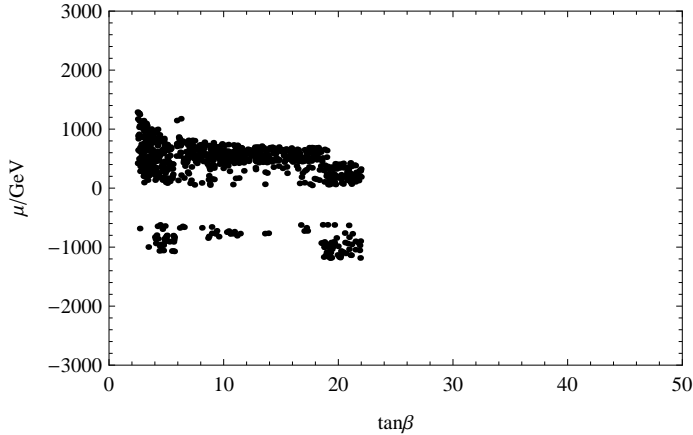}
\vspace{0cm}
\caption[]{The allowed parameters in the plane of $\tan\beta$ and $\mu$.}\label{TBMU}
\end{figure}

In order to scan the parameters more efficiently, we plot the parameters in $3\sigma$ sensitivity of the relic density with two variables.
With the parameters $M_1=1.2{\rm TeV},~ M_2=1{\rm TeV},~ M_{BL}=1{\rm TeV},~ M_{BB^\prime}=0.4{\rm TeV}$, we plot the
allowed results in the plane of $\tan\beta$ and $\mu$, which is shown in the Fig.\ref{TBMU}.
 $\tan\beta$ appears in almost all the mass matrixes of Fermions, scalars and Majoranas, and it must be a sensitive parameter. From the
 Fig. \ref{TBMU}, one can find that $\tan\beta$ should be in the region from 2.5 to 23.
 The corresponding values of $\mu$ are approximately in the scope $(-1100\sim-600)$ GeV and $(0\sim 1100)$ GeV. The allowed region of $\mu>0$
 is larger than that of $\mu<0$.

\begin{figure}[h]
\setlength{\unitlength}{1mm}
\centering
\hspace{1.3cm}\includegraphics[width=3.6in]{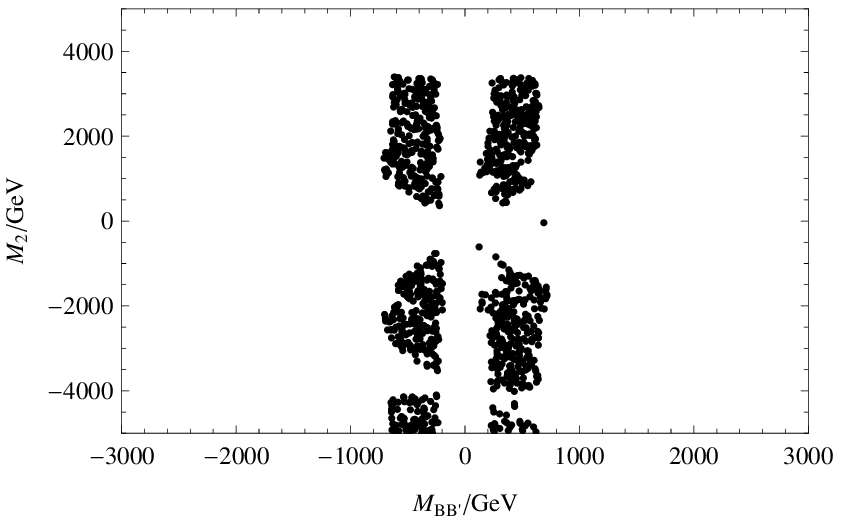}
\vspace{0cm}
\caption[]{The allowed parameters in the plane of $M_{BB^\prime}$ and $M_2$.}\label{MBBPM2}
\end{figure}

\begin{figure}[h]
\setlength{\unitlength}{1mm}
\centering
\hspace{1.3cm}\includegraphics[width=3.6in]{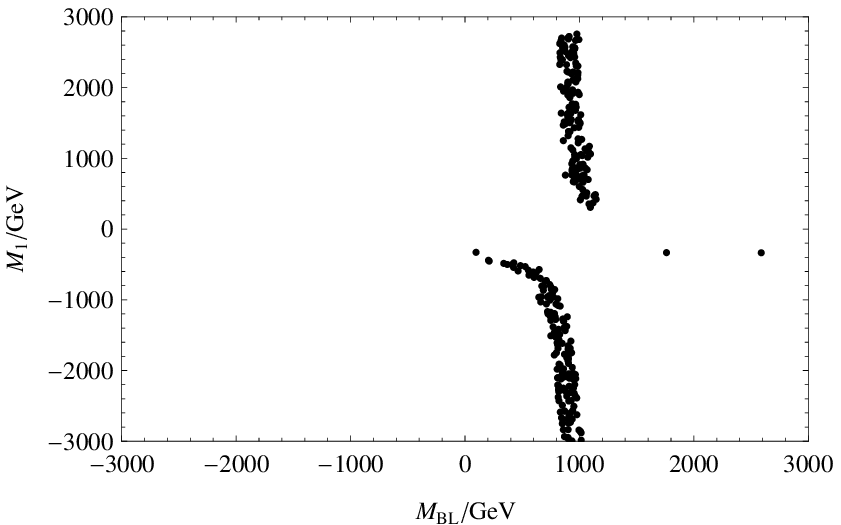}
\vspace{0cm}
\caption[]{The allowed parameters in the plane of $M_{BL}$ and $M_1$.}\label{MBLM1}
\end{figure}

$M_{BB^\prime}$, $M_2$, $M_{BL}$ and $M_1$ are all in the mass matrix of neutralino.
 So we study their effects and allowed regions.
 In the Fig.\ref{MBBPM2}, we plot the results versus $M_{BB^\prime}$ and $M_2$ with $\tan\beta=9, \mu=0.5 {\rm TeV}, M_1=1.2{\rm TeV},  M_{BL}=1{\rm TeV}$.
$M_2$ smaller than 3500 GeV is acceptable. $M_{BB^\prime}$'s region is almost symmetric relative to $M_{BB^\prime}=0$,
whose value should be in the region $(-700\sim-200)$ GeV and $(100\sim 600)$ GeV. The region near the point (0, 0) is excluded.
In the plane of $M_{BL}$ and $M_1$,
the numerical results of the relic density are researched as $\tan\beta=9, \mu=0.5 {\rm TeV},M_2=1{\rm TeV},  M_{BB^\prime}=0.4{\rm TeV}$.
It is obvious that the allowed region in the Fig.\ref{MBLM1} is smaller than that in the Fig.\ref{TBMU} and Fig.\ref{MBBPM2}.
The points gather around the narrow band near $M_{BL}=1000$ GeV.

\begin{figure}[h]
\setlength{\unitlength}{1mm}
\centering
\hspace{1.3cm}\includegraphics[width=3.4in]{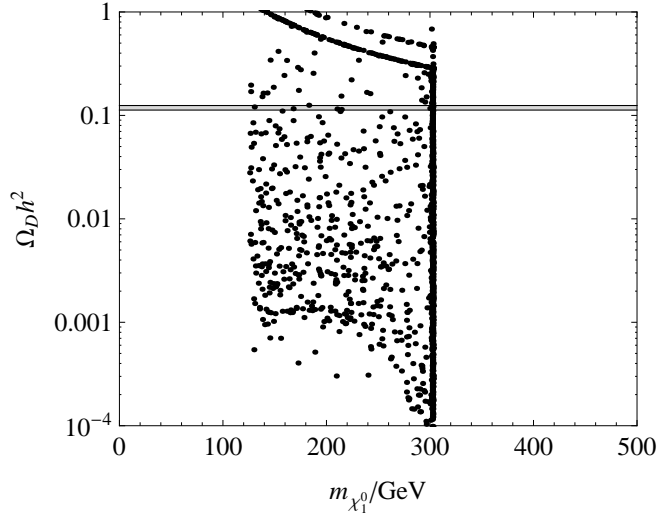}
\vspace{0cm}
\caption[]{The relic density $\Omega_D h^2$ versus the lightest neutralino mass $M_{\chi^0_1}$}\label{FDMD}
\end{figure}

To find large parameter space satisfying the relic density, we plot the relic density versus the lightest neutralino mass $M_{\chi^0_1}$
with the parameters: $2\leq\tan\beta\leq50$, and $-2 {\rm TeV} \leq \mu\leq 2 {\rm TeV}$,
 $-2 {\rm TeV} \leq M_{BL}\leq 2 {\rm TeV}$, $-2 {\rm TeV} \leq M_{BB^\prime}\leq 2 {\rm TeV}$, $-2 {\rm TeV} \leq M_{1}\leq 2 {\rm TeV}$
 $-2 {\rm TeV} \leq M_{2}\leq 2 {\rm TeV}$. The results are shown in the Fig.\ref{FDMD}, where the gray band represents the relic density
 in three $\sigma$ sensitivity. One can easily see that large reasonable parameter space in near $M_{\chi^0_1}\sim 300{\rm GeV}$.
 In the $M_{\chi^0_1}$ region (120 GeV to 280 GeV), there are also reasonable parameter space for $\Omega_D h^2$, but these parameter space
 are much smaller than the reasonable parameter space for $M_{\chi^0_1}\sim 300 {\rm GeV}$.

\subsection{The cross section of neutralino scattering off nucleus}

In this subsection, the cross section of the lightest neutralino scattering off nucleus is numerically researched
with the parameters $M_1 =1.2 {\rm TeV},~M_2 =1 {\rm TeV},~ M_{BL}=1 {\rm TeV},~\tan\beta =9,~\mu=0.5{\rm TeV},
~ M_{BB^\prime}=0.4{\rm TeV}$. Both the spin-independent cross section and spin-dependent cross section are studied here.
The constraint from the relic density is taken into account.
 In our used parameter space, the mass of the lightest neutralino is about 300 GeV.
For dark matter mass $\sim$ 300 GeV,
the corresponding experimental limit on spin-independent direct detection is about $2.3\times10^{-46}~{\rm cm}^2$ for Xenon in $1\sigma$ sensitivity.
While, it is about twice as large for PandaX \cite{PanXen1,PanXen2}. The experimental constraint on spin-independent cross section
is much more severe than that on spin-dependent cross section. The  direct detection experimental limit on spin-dependent cross section
is about $4.0\times10^{-41}~{\rm cm}^2$ for Xenon1T  experiment. The corresponding constraint is around $1.4\times10^{-40}~{\rm cm}^2$
for PandaX-II\cite{DarkSUSY1,PanXen1}.

\begin{figure}[h]
\setlength{\unitlength}{1mm}
\centering
\hspace{1.3cm}\includegraphics[width=3.6in]{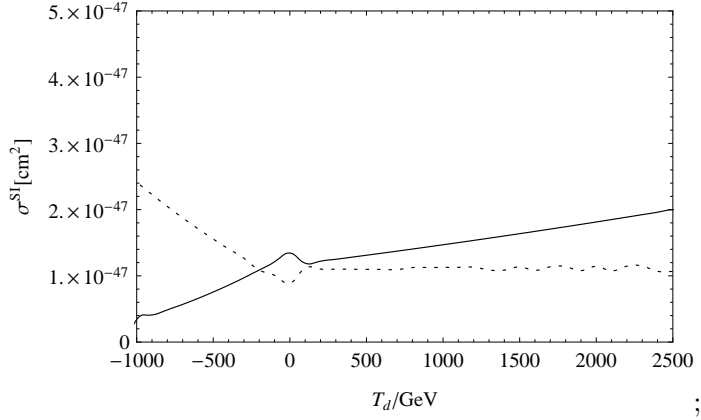};
\vspace{0cm}
\caption[]{The spin-independent cross section versus $T_d$, the solid line corresponds to $M_D^2$=6${\rm TeV}^2$ and the
dotted line corresponds to $M_D^2$=5 ${\rm TeV}^2$.}\label{Td}
\end{figure}

To simplify the discussion, we suppose $M^2_{Q11}=M^2_{Q22}=M_Q^2$. At first,
the spin-independent cross section is researched with the parameters $M_Q^2=$ 10 ${\rm TeV}^2$.
$T_d$ is in the non-diagonal element of the mass squared matrix for scalar down type quarks.
Therefore, $T_d$ should influence the scattering cross section.
In the Fig.\ref{Td}, the numerical results of the spin-independent cross section versus $T_d$ are plotted by the
solid line($M_D^2=6 {\rm TeV}^2$) and dotted line ($M_D^2=5 {\rm TeV}^2$).
    Generally speaking, the both lines are at the order of  $10^{-47}~{\rm cm}^2$, which are about
one order smaller than the experimental bound.

\begin{figure}[h]
\setlength{\unitlength}{1mm}
\centering
\hspace{1.3cm}\includegraphics[width=3.6in]{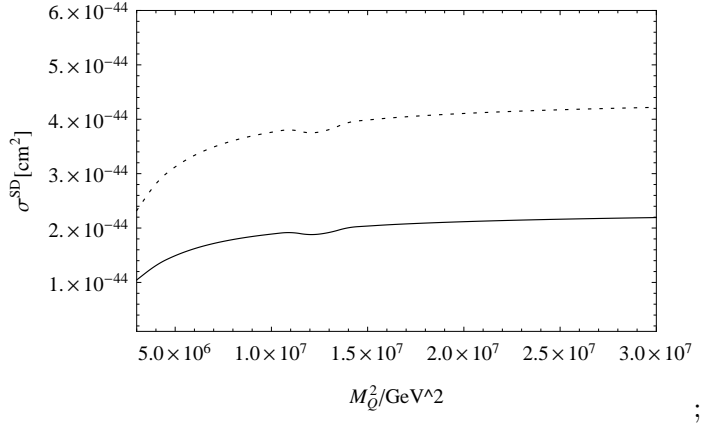};
\vspace{0cm}
\caption[]{The spin-dependent cross section versus $M_{Q}^2$ is plotted by the solid line (dotted line) with $M_D^2=10(5) $${\rm TeV}^2$.}\label{MQ2}
\end{figure}

Secondly,  we calculate the spin-dependent cross section as $T_d=1 {\rm TeV}$. $M_Q^2$ are the important diagonal
elements in the mass squared matrixes of scalar quarks, and they can strongly affect the masses of scalar quarks.
In the Fig.\ref{MQ2}, the numerical results of spin-dependent cross section versus $M_Q^2$ are represented by the solid
 line (dotted line) with $M_D^2 =10 {\rm TeV}^2~(5 {\rm TeV}^2)$. The dotted line is above the solid line. With
the same $M_{Q}^2$ in the region($3.0\times10^6 {\rm GeV}^2\sim3.0\times10^7 {\rm GeV}^2$), the values of the dotted line are about
$2\times10^{-44} {\rm cm}^2$ larger than the values of the solid line. Generally speaking, they are about three orders smaller than
the experimental bounds.

\section{discussion and conclusion}
We extend MSSM with the $U(1)_X$ local gauge group and obtain the so called $U(1)_X$SSM. In the $U(1)_X$SSM, there  are
several superfields beyond MSSM, such as right-handed neutrinos, three singlet Higgs superfields $\hat{\eta},~\hat{\bar{\eta}},~\hat{S}$.
 As discussed in MSSM, the lightest neutralino is studied in detail as dark matter candidate.
While,  both the lightest sneutrino and the lightest neutralino can be dark matter candidates in $U(1)_X$SSM.
Supposing the lightest CP-even sneutrino as LSP and dark matter candidate, we research its relic density and the scattering cross
section off nucleus in our previous work\cite{ZSMJHEP}. $U(1)_X$SSM has richer phenomenology than MSSM. To compare the scalar neutrino condition, we research the lightest neutralino as dark matter candidate in this work.

To calculate the relic density of $\chi^0_1$, we consider the self-annihilation and co-annihilation processes. In our used parameter space,
the masses of the SUSY particles except $\chi^0_1$ are all heavier enough than the mass of $\chi_1^0$. Therefore, self-annihilation processes
are dominant and co-annihilation processes are suppressed by the exponential function. In the whole, this is the general condition. The resonance annihilation does not take place, because the masses of the exchanged virtual particles are not near $2*m_{\chi^0_1}$. From our numerical results, we find that $M_{BB^\prime}$ and $M_{BL}$ in the neutralino mass matrix are
sensitive parameters for the relic density. The reason is that both $M_{BB^\prime}$ and $M_{BL}$ affect neutralino mixing.
Large reasonable parameter space supports $m_{\chi_1^0}\sim 300 {\rm GeV}$, though $m_{\chi_1^0}$ can be
smaller with reasonable parameter space.
The obtained numerical results can well satisfy the experimental constraints from the relic density of dark matter.
The cross section of $\chi_1^0$ scattering off nucleus are also calculated in this work. The spin-independent and spin-dependent
cross sections are at least one order smaller than their experimental constraints. This work makes up for the dark matter research\cite{ZSMJHEP}, where just the lightest CP-even scalar neutrino is supposed as dark matter.

{\bf Acknowledgments}

This work is supported by National Natural Science Foundation of China (NNSFC)
(No. 11535002, No. 11705045, No. 11605037), Natural Science Foundation of Hebei Province
(A2020201002), Post-graduate's Innovation Fund Project of Hebei Province
 (No. CXZZBS2019027),  and the youth top-notch talent support program of the Hebei Province.

\appendix

\section{The coupling}

The couplings of neutralino and gauge bosons are $\chi^0-\chi^0-Z$, $\chi^0-\chi^0-Z^\prime$ and $\chi^0-\chi^\pm-W^\pm$. Their
concrete forms are shown as
\begin{eqnarray}
&&\mathcal{L}_{\chi^0\chi^0Z}=\bar{\chi}^0_i\Big\{-\frac{i}{2} \Big[ \Big(g_1 \cos{\theta'}_W  \sin\theta_W
 + g_2 \cos\theta_W  \cos{\theta'}_W   - (g_{Y X} + g_{X})\sin{\theta'}_W  \Big) \nonumber \\&&\hspace{1.7cm}\times(N^*_{j 3}N_{{i 3}}
 - N^*_{j 4} N_{{i 4}}) - 2g_{X} \sin{\theta'}_W \Big(N^*_{j 6} N_{{i 6}}  - N^*_{j 7} N_{{i 7}} \Big)\Big]\gamma_{\mu}P_L\nonumber\\
  &&\hspace{1.7cm} +\frac{i}{2} \Big[\Big(g_1 \cos{\theta'}_W  \sin\theta_W   + g_2 \cos\theta_W  \cos{\theta'}_W   - (g_{Y X} + g_{X})\sin{\theta'}_W  \Big)\nonumber \\ &&\hspace{1.7cm}\times(N^*_{i 3}N_{{j 3}}
 - N^*_{i 4} N_{{j 4}})- 2g_{X} \sin{\theta'}_W(N^*_{i 6} N_{{j 6}}  - N^*_{i 7} N_{{j 7}})\Big]\gamma_{\mu}P_R\Big\}\chi^0_jZ^\mu,\\
 &&\mathcal{L}_{\chi^0\chi^0Z^\prime}=\bar{\chi}^0_i\Big\{\frac{i}{2} \Big(
 [(g_1 \sin\theta_W   + g_2 \cos\theta_W )\sin{\theta'}_W   + (g_{Y X} + g_{X})\cos{\theta'}_W ] \nonumber \\
 &&\hspace{1.7cm}\times(N^*_{j 3}N_{{i 3}}-N^*_{j 4}N_{{i 4}})
 +2g_{X} \cos{\theta'}_W(N^*_{j 6} N_{{i 6}}  - N^*_{j 7} N_{{i 7}})\Big)\gamma_{\mu}P_L\nonumber\\
  && \hspace{1.7cm}-\frac{i}{2} \Big( [(g_1\sin\theta_W   + g_2 \cos\theta_W)\sin{\theta'}_W
   + (g_{Y X} + g_{X})\cos{\theta'}_W]\nonumber \\ &&\hspace{1.7cm}\times(N^*_{i 3}N_{{j 3}}-N^*_{i 4}N_{{j 4}})  +2 g_{X} \cos{\theta'}_W
   (N^*_{i 6} N_{{j 6}}  - N^*_{i 7} N_{{j 7}} )\Big)\gamma_{\mu}P_R\Big\}\chi^0_jZ^{\prime\mu},\\
 &&\mathcal{L}_{\chi^0\chi^\pm W}=-\frac{i}{2}\bar{\chi}^+_i\Big\{ g_2 (2 N^*_{j 2} U_{{i 1}}  + \sqrt{2} N^*_{j 3} U_{{i 2}} )\gamma_{\mu}P_L\nonumber\\&&\hspace{1.9cm} + g_2 (2 V^*_{i 1} N_{{j 2}}  - \sqrt{2} V^*_{i 2} N_{{j 4}} )\gamma_{\mu}P_R\Big\}\chi^0_jW^{+\mu}.
  \end{eqnarray}
  $U$ and $V$ are the rotation matrixes to diagonalize chargino mass matrix.
The couplings $\chi^0-\chi^0-Z$ and $\chi^0-\chi^0-Z^\prime$ contribute to the self-annihilation, and the coupling $\chi^0-\chi^\pm-W^\pm$ gives correction to co-annihilation. We deduce the coupling of neutralino-lepton-slepton($\chi^0-l-\tilde{L}$).
\begin{eqnarray}
 &&\mathcal{L}_{\chi^0l\tilde{L}}=\bar{\chi}^0_i\Big\{i \Big(\frac{1}{\sqrt{2}}( g_1 N^*_{i 1}
  + g_2 N^*_{i 2}   +g_{Y X} N^*_{i 5} )Z_{{k j}}^{E} - N^*_{i 3} Y_{e,j} Z_{{k 3 + j}}^{E}   \Big)P_L\nonumber\\
  &&\hspace{1.5cm} - i\Big(\frac{1}{\sqrt{2}} Z_{{k 3 + j}}^{E} [2 g_1 N_{{i 1}}  + (2 g_{Y X}  + g_{X})N_{{i 5}} ]
  +Y^*_{e,j}  Z_{{k j}}^{E}  N_{{i 3}} \Big)P_R\Big\}e_j\tilde{L}_k.
  \end{eqnarray}
  $Z^E$ is used to diagonalize the mass squared matrix of slepton.

The coupling of neutralino-neutralino-CP-even Higgs($\chi^0-\chi^0-H$) is
\begin{eqnarray}
 &&\mathcal{L}_{\chi^0\chi^0H}=\bar{\chi}^0_i\Big\{\frac{i}{2} \Big[\Big(2 g_{X} (N^*_{i 6} N^*_{j 5}
 + N^*_{i 5} N^*_{j 6})  - \sqrt{2} {\lambda}_{C} (N^*_{i 8} N^*_{j 7}+ N^*_{i 7} N^*_{j 8}) \Big)Z_{{k 3}}^{H}\nonumber \\&&+\Big(N^*_{i 3}[g_1 N^*_{j 1} - g_2 N^*_{j 2}
 +(g_{Y X}  +g_{X} )N^*_{j 5}]+N^*_{j 3}[g_1 N^*_{i 1} - g_2 N^*_{i 2}  +(g_{Y X}  +g_{X} )N^*_{i 5}]\nonumber\\&&
   +\sqrt{2} {\lambda}_{H} (N^*_{i 8} N^*_{j 4}+ N^*_{i 4} N^*_{j 8})\Big)Z_{{k 1}}^{H} +\Big(N^*_{i 4}(g_2  N^*_{j 2}- g_{Y X} N^*_{j 5}
    - g_{X}  N^*_{j 5}- g_1  N^*_{j 1})\nonumber \\ &&
  + N^*_{j 4}(g_2 N^*_{i 2} - g_{Y X} N^*_{i 5}  - g_{X} N^*_{i 5}
   - g_1N^*_{i 1} )+\sqrt{2} {\lambda}_{H} (N^*_{i 8} N^*_{j 3}
+N^*_{i 3} N^*_{j 8})\Big) Z_{{k 2}}^{H}\nonumber \\
& &+\Big(\sqrt{2} {\lambda}_{H} (N^*_{i 4} N^*_{j 3} + N^*_{i 3}N^*_{j 4})-2 \sqrt{2} \kappa N^*_{i 8} N^*_{j 8}- \sqrt{2} {\lambda}_{C} (N^*_{i 7} N^*_{j 6}
 + N^*_{i 6} N^*_{j 7})\Big) Z_{{k 5}}^{H}
   \nonumber \\
& &-\Big(2 g_{X}( N^*_{i 7} N^*_{j 5} + N^*_{i 5} N^*_{j 7})+ \sqrt{2} {\lambda}_{C} (N^*_{i 8} N^*_{j 6}+ N^*_{i 6} N^*_{j 8})\Big)Z_{{k 4}}^{H}
\Big]P_L\nonumber\\
 & & + \,\frac{i}{2} \Big[Z_{{k 5}}^{H}\Big(
 \sqrt{2} {\lambda}_{H}^* (N_{{i 4}} N_{{j 3}}
  + N_{{i 3}} N_{{j 4}})- \sqrt{2} {\lambda}_{C}^* ( N_{{i 7}} N_{{j 6}} + N_{{i 6}} N_{{j 7}}) -2 \sqrt{2} \kappa^*  N_{{i 8}} N_{{j 8}}\Big) \nonumber \\
& &+ \Big(N_{{j 3}}[g_1N_{{i 1}} - g_2 N_{{i 2}}
+(g_{Y X} +g_{X} )N_{{i 5}}]
 +N_{{i 3}}[g_1 N_{{j 1}}  - g_2 N_{{j 2}}
+ (g_{Y X} + g_{X})N_{{j 5}}]\nonumber \\
& &+\sqrt{2} {\lambda}_{H}^* (N_{{i 8}} N_{{j 4}} + N_{{i 4}} N_{{j 8}}) \Big)Z_{{k 1}}^{H}
- Z_{{k 2}}^{H} \Big([g_1 N_{{i 1}}  - g_2 N_{{i 2}}  + (g_{Y X} + g_{X})N_{{i 5}} ]N_{{j 4}}
 \nonumber \\&&+N_{{i 4}} [g_1N_{{j 1}}  - g_2 N_{{j 2}}  + (g_{Y X} + g_{X})N_{{j 5}} ]
 - \sqrt{2} {\lambda}_{H}^* (N_{{i 3}} N_{{j 8}}  + N_{{i 8}} N_{{j 3}})\Big)\nonumber \\
& &- Z_{{k 4}}^{H}\Big(\sqrt{2} {\lambda}_{C}^*  (N_{{i 8}} N_{{j 6}}+ N_{{i 6}} N_{{j 8}})+2 g_{X} (N_{{i 7}} N_{{j 5}} +  N_{{i 5}} N_{{j 7}})\Big) \nonumber \\
& &+Z_{{k 3}}^{H}\Big(2 g_{X} (N_{{i 5}} N_{{j 6}}+ N_{{i 6}} N_{{j 5}})
 - \sqrt{2} {\lambda}_{C}^*  (N_{{i 8}} N_{{j 7}} + N_{{i 7}} N_{{j 8}})\Big)\Big]P_R\Big\}\chi^0_jH_k
 \end{eqnarray}
  The concrete form of neutalino-chargino-charged Higgs coupling ($\chi^0-\chi^\pm-H^\pm$) is
 \begin{eqnarray}
 &&\mathcal{L}_{\chi^0\chi^\pm H^\pm}=\bar{\chi}^+_i\Big\{\frac{i}{2} \Big[- V^*_{i 2} \Big(2 {\lambda}_{H} N^*_{j 8} Z_{{k 1}}^{+}  + \sqrt{2} [g_1   + g_2 N^*_{j 2}  + (g_{Y X} + g_{X})N^*_{j 5} ]Z_{{k 2}}^{+} \Big)\nonumber\\  &&\hspace{1.8cm}-2 g_2 V^*_{i 1} N^*_{j 4} Z_{{k 2}}^{+}  \Big]P_L +\frac{i}{2} \Big[-2 g_2 U_{{i 1}} N_{{j 3}} Z_{{k 1}}^{+} +U_{{i 2}} \Big(-2 {\lambda}_{H}^* N_{{j 8}} Z_{{k 2}}^{+} \nonumber\\  &&\hspace{1.8cm} + \sqrt{2}[ g_1N_{{j 1}}   +  g_2 N_{{j 2}}   +  (g_{X} +  g_{Y X} )N_{{j 5}} ]Z_{{k 1}}^{+} \Big)\Big]P_R\Big\}\chi^0_jH^+_k.
 \end{eqnarray}
Neutralinos interact with neutrinos and sneutrinos in the following form
 \begin{eqnarray}
 &&\mathcal{L}_{\chi^0\nu\tilde{\nu^I}}=\bar{\chi}^0_i\Big\{\frac{1}{2} \sum_{a=1}^{3}\Big(- \sqrt{2} N^*_{i 7} \sum_{b=1}^{3}Y_{x,{a b}}  (Z^{I,*}_{k 3 + b}U^{V,*}_{j 3 + a}
+U^{V,*}_{j 3 + b} Z^{I,*}_{k 3 + a} )\nonumber \\
 &&\hspace{1.5cm}+( g_{Y X} N^*_{i 5} - g_2 N^*_{i 2}
 + N^*_{i 1}  g_1  )U^{V,*}_{j a}Z^{I,*}_{k a} - g_{X} N^*_{i 5}U^{V,*}_{j 3 + a} Z^{I,*}_{k 3 + a}     \Big)P_L\nonumber\\
  && \hspace{1.5cm} +\frac{1}{2} \sum_{a=1}^{3}\Big( \sqrt{2}\sum_{b=1}^{3}
 \Big(Z^{I,*}_{k 3 + b} U_{{j 3 + a}}^{V}   + Z^{I,*}_{k 3 + a}  U_{{j 3 + b}}^{V} \Big)Y^*_{x,{a b}}N_{{i 7}}\nonumber \\
 &&\hspace{1.5cm}+Z^{I,*}_{k 3 + a} U_{{j 3 + a}}^{V} g_{X} N_{{i 5}}- Z^{I,*}_{k a} U_{{j a}}^{V}  \Big(g_1 N_{{i 1}}
  - g_2 N_{{i 2}}  + g_{Y X} N_{{i 5}} \Big)  \Big)P_R\Big\}\nu_j\tilde{\nu}^I_k,
    \end{eqnarray}
 There are also neutralino-quark-squark couplings
   \begin{eqnarray}
 &&\mathcal{L}_{\chi^0d\tilde{D}}=-\frac{i}{6}\bar{\chi}^0_i\Big\{\Big(\sqrt{2}( g_1 N^*_{i 1}
 -3 g_2 N^*_{i 2}   + g_{Y X} N^*_{i 5}) Z_{{k j}}^{D} +6 N^*_{i 3} Y_{d,j} Z_{{k 3 + j}}^{D}   \Big)P_L\nonumber\\
  && \hspace{1.6cm}+\Big(6 Y^*_{d,j}   Z_{{k j}}^{D}  N_{{i 3}}
  + \sqrt{2} Z_{{k 3 + j}}^{D}  [2 g_1 N_{{i 1}}  + (2 g_{Y X}  + 3 g_{X})N_{{i 5}}]\Big)P_R\Big\}d_j\tilde{D}^*_k,
 \\
 &&\mathcal{L}_{\chi^0u\tilde{U}}=-\frac{i}{6}\bar{\chi}^0_i\Big\{ \Big(\sqrt{2}( g_1 N^*_{i 1}
   +3  g_2 N^*_{i 2}   + g_{Y X} N^*_{i 5}   ) Z_{{k j}}^{U}+6 N^*_{i 4} Y_{u,j} Z_{{k 3 + j}}^{U}   \Big)P_L\nonumber\\
  &&\hspace{1.5cm}- \Big( \sqrt{2}Z_{{k 3 + j}}^{U}   \Big((3 g_{X}  + 4 g_{Y X})N_{{i 5}}  + 4 g_1 N_{{i 1}} -6 Y^*_{u,j}   Z_{{k j}}^{U}  N_{{i 4}} \Big)\Big)P_R\Big\}u_j\tilde{U}^*_k.
  \end{eqnarray}

\end{document}